\documentclass[journal,comsoc]{IEEEtran}

\usepackage[caption=false,font=normalsize,labelfont=sf,textfont=sf]{subfig}
\usepackage{algorithm}
\usepackage{algpseudocode}
\usepackage{amsmath}
\usepackage{float}
\usepackage{graphicx,dblfloatfix}
\usepackage{mathtools}
\DeclarePairedDelimiter{\ceil}{\lceil}{\rceil}
\DeclareMathOperator*{\argmax}{arg\,max}
\usepackage{relsize}

\usepackage{algpseudocode}

\begin{document}
\title{\LARGE{Throughput Optimized Non-Contiguous Wideband Spectrum Sensing via Online Learning and Sub-Nyquist Sampling}}

\author{Himani~Joshi,~%~\IEEEmembership{Member,~IEEE,}
	Sumit J Darak,~%~\IEEEmembership{Fellow,~OSA,}
	A Anil Kumar~%~\IEEEmembership{Fellow,~OSA,}
	and~Rohit~Kumar%~\IEEEmembership{Life~Fellow,~IEEE}% <-this % stops a space
	\thanks{Himani Joshi and Sumit J Darak are with ECE Department, IIIT-Delhi, India e-mail: \{himanij,sumit\}@iiitd.ac.in,~A Anil Kumar is with TCS Labs, Bangalore, India email:~achannaanil.kumar@tcs.com,~Rohit Kumar is with NIT-Delhi, India email:~rohitkumar@nitdelhi.ac.in}}

%\author{Himani~Joshi,~%~\IEEEmembership{Member,~IEEE,}
%	Sumit J Darak,~%~\IEEEmembership{Fellow,~OSA,}
%	A Anil Kumar,~%~\IEEEmembership{Fellow,~OSA,}
%	and~Rohit~Kumar%~\IEEEmembership{Life~Fellow,~IEEE}% <-this % stops a space
%	\thanks{Himani Joshi and Sumit J Darak are IIIT-Delhi, India-110020 e-mail: \{himanij,sumit\}@iiitd.ac.in}% <-this % stops a space
%	\thanks{A Anil Kumar is with TCS Labs, Bangalore, India-560012 email:achannaanil.kumar@tcs.com}% <-this % stops a space
%	\thanks{Rohit Kumar is with NIT-Delhi, India-110040 email:rohitkumar@nitdelhi.ac.in}}

%	\author{\IEEEauthorblockN{Himani~Joshi\IEEEauthorrefmark{1},
%			Sumit~J~Darak\IEEEauthorrefmark{1}
%			A Anil Kumar\IEEEauthorrefmark{2} and Rohit Kumar\IEEEauthorrefmark{3}}\\%\IEEEauthorieeemembermark{1}}
%		\IEEEauthorblockA{\IEEEauthorrefmark{1} ECE Dept, IIIT-Delhi
%%			Department of Electronics and Communication Engineering, IIIT-Delhi, India-110020\\
%			\IEEEauthorrefmark{2} TCS Lab, Bangalore
%%			TCS Innovations Lab, Bangalore, India-560012\\
%			\IEEEauthorrefmark{3} ECE Dept, NIT-Delhi 
%			%Departement of Electronics and Communication Engineering, NIT-Delhi, India -110040\\
%			}
%}	

%\author{Author 1, Author 2, Author 3, Author 4}
\maketitle
\begin{abstract}		
		In this paper, we consider non-contiguous wideband spectrum sensing (WSS) for spectrum characterization and allocation in next generation heterogeneous networks. The proposed WSS consists of sub-Nyquist sampling and digital reconstruction to sense multiple non-contiguous frequency bands. Since the throughput (i.e. the number of vacant bands) increases while the probability of successful reconstruction decreases with increase in the number of sensed bands, we develop an online learning algorithm to characterize and select frequency bands based on their spectrum statistics. We guarantee that the proposed algorithm allows sensing of maximum possible number of frequency bands and hence, it is referred to as throughput optimized WSS. We also provide a lower bound on the number of time slots required to characterize spectrum statistics. Simulation and experimental results in the real radio environment show that the performance of the proposed approach converges to that of Myopic approach which has prior knowledge of spectrum statistics.
\end{abstract}
\vspace{-0.1cm}
\vspace{-0.2cm}
\section{Introduction}
\vspace{-0.15cm}
Next generation wireless networks are envisioned on a revolutionary path of spectrum sharing to support a wide range of deployments and services from enhanced mobile broadband to ultra-reliable low-latency communications.
%to massive machine type communications. 
The 3GPP new radio (NR) is expected to operate not only in the licensed spectrum but also in the shared (2.3 GHz Europe / 3.5 GHz USA) as well as unlicensed spectrum (2.4 GHz / 5-7 GHz / 57-71 GHz global). Hence, base stations or geolocation database need wideband spectrum sensing (WSS) to sense a spectrum of few tens of Gigahertz and dynamically allocate the desired spectrum to NRs~\cite{wss}. Recently, various sub-Nyquist sampling (SNS) and digital reconstruction based contiguous WSS methods have been proposed and they employ low rate analog-to-digital converters (ADCs) based on spectrum sparsity~\cite{sns}. In the wideband spectrum, some frequency bands are reserved for applications such as WiFi, military and radar systems while some are crowded or may not be useful. Hence, non-contiguous WSS needs to be explored as it offers complete control over the number and location of bands thereby significantly increases the sensing bandwidth.

		To the best of our knowledge, \cite{mab_cs} is the only work which has explored the non-contiguous WSS approach. However, \cite{mab_cs} assumes the complete knowledge of spectrum statistics which makes the frequency band selection trivial. Since the  throughput (i.e. the number of vacant bands) increases while the probability of successful reconstruction decreases with an increase in the number of selected bands, we develop an online learning algorithm which based on the spectrum statistics characterizes and determines frequency bands for sensing.
		%we develop an online learning algorithm to characterize and select frequency bands based on their spectrum statistics.
		 We also provide theoretical guarantees, extensive simulation and experimental results in the real radio environment to validate the superiority of the proposed approach. We begin with the signal model in the next section.	
%		 In this paper, we develop a sequential online learning algorithm to characterize various sections of the wideband spectrum based on their spectrum statistics and simultaneously select multiple frequency bands via SNS. We guarantee that the proposed approach allows sensing of maximum possible number of frequency bands for a given number of ADCs, and hence, it is referred to as throughput optimised WSS. Here, throughput is directly proportional to the number of vacant frequency bands identified in a given time slot. A lower bound on the number of time slots required to learn the spectrum statistics is also derived. To further validate the performance of the proposed algorithm on real radio environment, a universal software radio peripheral (USRP) testbed is also developed. We begin with the signal model in the next section.
\vspace{-0.25cm}
\section{Signal Model}	
\vspace{-0.15cm}
		We consider a wideband signal $x(t)$ of bandwidth $f_{max}$. It is divided into $N$ frequency bands which evolve as independent two states (vacant and busy) Markovian chain and is given as 
		\vspace{-0.35cm}
		\begin{equation}
		x(t) = \sum_{i\in 1}^{I} a_i(t)e^{j2\pi f_it}
		\vspace{-0.2cm}
		\end{equation}	
		where $ a_i(t) $ is a narrowband signal transmitted at a carrier frequency $ f_i $ where $I\leq N$.
		Similar to \cite{mab_cs, q_lear_1}, assumptions made for the wideband signal, $x(t)$ are:
		\begin{enumerate}
			\item The divided $N$ frequency bands are static for a time slot, $t_s$ and have the uniform bandwidth, $B = \frac{f_{max}}{N}$.
			\item The bandwidth of all $a_i(t)$ can not exceed $B$ and they are orthogonal to each other, i.e. $ \{A_i(f)\cap A_j(f)\}=\emptyset \forall i\neq j$, where $ A_i(f) $ is the Fourier transform of~$ a_i(t) $.
		\end{enumerate}
		Consider a binary support vector, $ \textbf{s} = \left[s_n(t_s)\right] _{n=1}^{N} $ where 
		$ s_n(t_s) = 0$ (or $1$) implies that the $ n^{th} $ frequency~band is vacant (or busy) at time slot $t_s$. The status of $ n^{th} $ band evolves with a transition probability, $ p_{uv}^n = \mathbb{P}(s_n(t_s)=v | s_n(t_s-1)=u) $ where $ u,v \in \{0,1\} $ and $ \mathbb{P}(.) $ is a probability operator.
\vspace{-0.1cm}

\section{Proposed Work}
%\vspace{-0.1cm}
		\begin{figure}[b]
			\label{BD}	
			\vspace{-0.5em}
			\centering
			\includegraphics[scale=0.4]{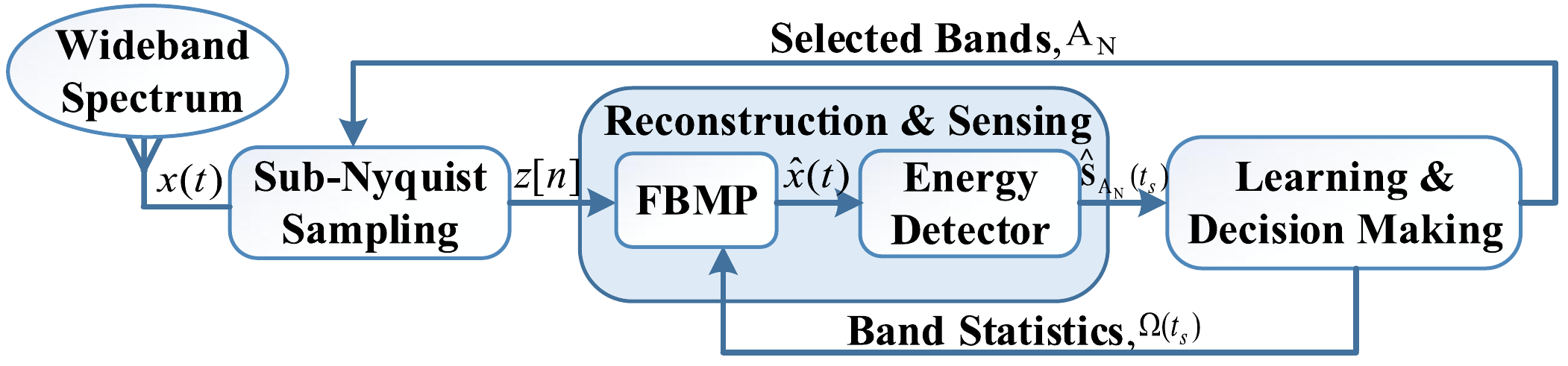}
			\vspace{-1em}
			\caption{Block diagram of the proposed WSS approach}
			\vspace{-1em}
		\end{figure}
		
			The proposed WSS approach consists of three blocks: 1)~SNS block, 2)~Reconstruction and sensing (RS) block, 3)~Learning and decision making (LDM) block. As shown in Fig.~1, the SNS block performs digitization on a~set~of~frequency bands, $A_N$  selected by the LDM block. The RS~block performs reconstruction of selected frequency bands from sub-Nyquist samples, $z[n]$ obtained from the SNS block followed by spectrum sensing to identify their vacant or busy status, $\textbf{s}_{A_N} \in \textbf{s}$. Based on $\textbf{s}_{A_N}$, the LDM block learns the frequency band statistics, $\Omega(t_s)$ via online learning algorithm and determines a set of frequency bands, $A_N$ for digitization in the subsequent time slot, $t_s$. Ideally, for a fixed number of ADC branches, $K$, as shown in Fig.~2, the LDM block should select as many most likely vacant bands as possible to maximize the throughput. However, due to increase in the number of sensed bands, $|A_N|$, the number of occupied bands in the sensed spectrum may increases which may lead to a decrease in the probability of successful reconstruction. Such a trade-off poses a real challenge in the design and integration of LDM block for spectrum sensing.% The detailed description on the designing of these blocks are presented in next sub-sections.

\vspace*{-\baselineskip}
\subsection{SNS Block}
The SNS block digitizes the non-contiguous frequency bands selected by the LDM block. Consider indexes of these bands are stored in a vector $A_N$ and cardinality $|A_N|$ indicates the number of selected bands. The SNS block, shown in Fig.~2, is based on the finite rate of innovation (FRI) technique \cite{fri} and consists of $K$ fixed parallel branches where the wideband signal, $ x(t) $ is passed through a branch dependent analog mixing function $p_k(t)$ given by  
\vspace{-0.5em}
\begin{equation}	  
p_k(t) = \sum_{n\in A_N}\alpha_{k,n}~ e^{-j2\pi f_nt}
\vspace{-0.2cm}
\end{equation}
\noindent where $ \alpha_{k,n} $ is a unique scaling coefficient for $n^{th}$ band~in~$A_N$ having center frequency $f_n$. The resultant signal is then~bandlimited by a low pass filter of bandwidth $B$ followed~by~digitization via ADCs of rate $\geq B$~Hz.
%	  The samples at the output of ADCs correspond to the chosen frequency bands indexed by $A_N$. 
The discrete time Fourier transform of samples generated at $k^{th}$ ADC is given by
\vspace{-0.5em}
\begin{equation}
Z_k(e^{j2\pi f/B}) = \sum_{n \in A_N} \alpha_{k,n} X_n(f+(n-1)B)~\forall f\in \left[0,B\right] 
\vspace{-0.2cm}
\end{equation}
where $ X_n(f) $ is the Fourier transform of the $n^{th}$ frequency band of $A_N$. The sub-Nyquist samples of all $ K $ ADCs can be collectively represented as
\vspace{-0.2cm}
\begin{equation}
\label{CS}
\textbf{Z}(e^{j2\pi f/B}) = \textbf{A} \textbf{X}_{A_N}(f)
\vspace{-0.3cm}
\end{equation}
where $ \textbf{X}_{A_N}(f) $ represents $ |A_N|\times 1 $ vector which contains Fourier transform of $ A_N $ frequency bands and $ \textbf{A} $ is a $ K \times |A_N| $  matrix containing $ \alpha_{k,n} $ as its $ (k,n)^{th} $ entry.   
\begin{figure}[b]	
	\centering	
	\vspace{-2em}
	\includegraphics[scale=0.28]{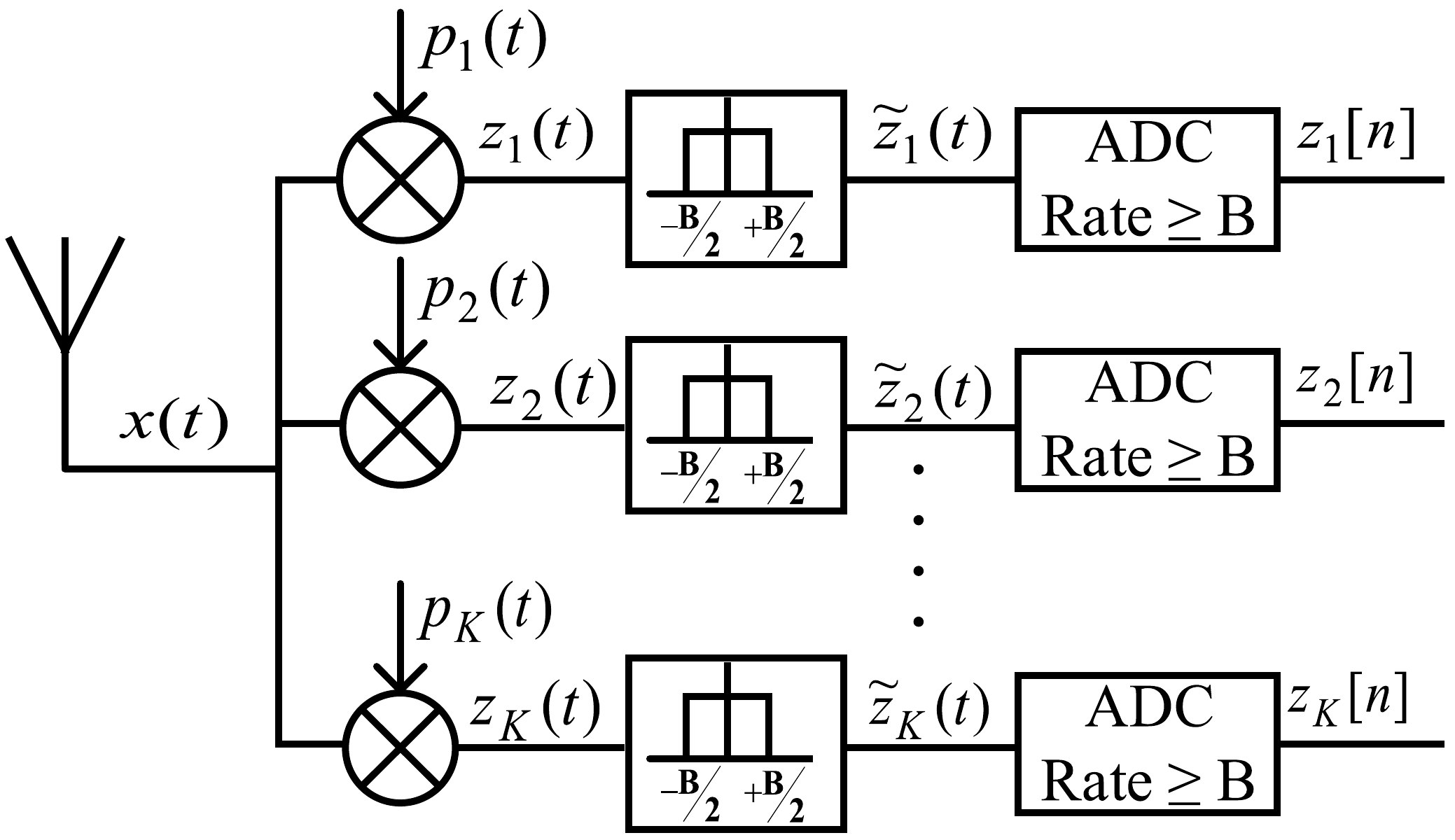}
	\vspace{-0.8em}
	\caption{Finite rate of innovation based SNS for multiband signal \cite{fri}.}
	%	\vspace{-1.9em}
\end{figure}
\vspace{-1em}
\subsection{RS Block}\label{rs}
\vspace{-0.1cm}
The RS block reconstructs $ A_N $ frequency bands from the sub-Nyquist samples, $ \textbf{Z} $ and then determines their status, $\textbf{s}_{A_N}$ via sensing technique.
For an error-free reconstruction of~$\textbf{X}_{A_{N}}(f)$~with~a fixed value of $K$, Eq.~\ref{CS} should satisfy the following two~criteria
\begin{enumerate}
	\item Kruskal rank of $\textbf{A}$, i.e. krank$(\textbf{A}) \geq K$. 
	To achieve this, we assume  $ \textbf{A} $  to be an independent and identically distributed Gaussian matrix. 
	\item For $|A_N|>K$, the number of busy frequency bands in $ A_N $ \cite{sns}, i.e. $|| \textbf{s}_{A_N} ||_0$, must be less than or equal to $ \Big\lfloor\frac{K}{2} \Big\rfloor $ 
%	where $ \textbf{s}_{A_N}\subset\textbf{s}$ denotes the support vector for $ A_N $ bands. 
\end{enumerate}

A Bayesian approach based reconstruction algorithms offer better reconstruction accuracy and lower complexity as compared to greedy and convex optimization based reconstruction algorithms \cite{fbmp}, respectively. Hence, we employ fast Bayesian matching pursuit (FBMP) reconstruction algorithm \cite{fbmp_2} which reconstructs $\textbf{X}_{A_N}$  by applying maximum a posteriori estimate to determine the best possible value of $ \textbf{s}_{A_N} $ for a given $ \textbf{Z} $. However, FBMP requires the prior knowledge of vacancy statistics of selected bands. In the proposed WSS approach, the LDM block aims to learn these statistics for frequency band selection and hence, they are readily available which makes FBMP a good fit for the proposed approach.
For sensing, we employ a simple energy detector to determine the status, $ \textbf{s}_{A_N} $ of $A_N$ bands based on their energy levels. However, the proposed approach can be extended to other detectors such as cyclostationary or Eigen value based detectors \cite{rec} which offer better performance but have higher computation complexity.

\vspace{-1em}
\subsection{LDM Block}
\label{P_LDM}
\vspace{-0.2em}
Based on the status, $\textbf{s}_{A_N}$ obtained from the RS block, the proposed LDM block performs three tasks: 1)~Characterization of frequency bands to estimate their statistics, 2)~Calculation of $|A_N|$ and 3)~Selection of frequency bands, $A_N$ for digitization by the SNS block. Consider a vector $ \mathbf{\Omega}(t_s) $ that is updated at every time slot and is defined as  
\vspace{-0.1cm}
\begin{equation}
\mathbf{\Omega}(t_s) = [\omega_1(t_s),\omega_2(t_s),.....,\omega_N(t_s)]
\vspace{-0.15cm}
\end{equation}
where $ \omega_n(t_s) = \mathbb{P}[s_n(t_s)=0]$ is an immediate probability of vacancy of $n^{th}$ band estimated at the time slot $t_s$. Now, based on the status, $ \textbf{s}_{A_N} $, the LDM block determines whether the reconstruction of $A_N$ bands is successful ($\xi_{A_N} = 0$) or not ($\xi_{A_N} = 1$). Depending on the reconstruction success, the value of $ \mathbf{\Omega}(t_s) $ is updated for the next time slot $t_s+1$ as
\vspace{-0.05cm}
\begin{equation}
\label{B_up}
{\omega_n(t_s+1) = \begin{cases}
\hat{p}_{10}^n, & \mbox{if}~ n\in A_N, s_n(t_s) = 1, \xi_{A_N} = 0\\
\hat{p}_{00}^n, & \mbox{if}~ n\in A_N, s_n(t_s) = 0, \xi_{A_N} = 0\\
\phi_n(t_s+1), & \mbox{if}~ n\notin A_N~ \mbox{or}~ \xi_{A_N} = 1\\
\end{cases}}
\vspace{-0.1cm}
\end{equation}
where $ \phi_n(t_s+1) = (1-\omega_n(t_s))\hat{p}_{10}^n+\omega_n(t_s)\hat{p}_{00}^n$ and $ \hat{p}_{uv}^n $ is the estimated transition probability. %Note that the transition probabilities are unknown and need to be estimated.

We consider two scenarios for $|A_N|$: 1)~$|A_N| = K $, and 2)~$|A_N| > K $. When $|A_N| = K $, i.e. the number of~ADCs,~the matrix $ \textbf{A} $ in Eq.~\ref{CS} becomes a full rank matrix which means that the system defined by Eq.~\ref{CS} exhibits a unique solution and hence, reconstruction is always successful ($\xi_{A_N} = 0$). However, when $ |A_N| > K $
and if $A_N$ does not meet the second requirement discussed in Section~\ref{rs}, then reconstruction fails. Mathematically,
\vspace{-0.2cm}
\begin{equation}
\xi_{A_N} = \begin{cases}
0 & \mbox{if} \quad ||\textbf{s}_{A_N}||_0 \leq \Gamma\\
1 & \mbox{otherwise}
\end{cases}
\vspace{-0.3cm}
\end{equation}
\vspace{0.15cm}
where
%\begin{equation*}
$\Gamma =  \begin{cases} |A_N| & \mbox{if} \quad |A_N|\leq K\\
\Big\lfloor\frac{K}{2}\Big\rfloor & \mbox{if} \quad K<|A_N|\leq N
\end{cases}
$

Since the transition probabilities are unknown, we propose LDM algorithm to estimate them and select $K$ most likely vacant bands out of $N$ bands in each time slot, i.e. $|A_N|= K $. Later, we discuss the optimized LDM algorithm which can select $|A_N|\geq K$ bands in each time slot.

%inspired from $ \epsilon $-greedy algorithm \cite{q_lear_1},
\subsubsection{LDM Using Online Learning}
The proposed LDM algorithm consists of two phases:~1)~Exploration phase to learn spectrum statistics of all $N$ bands and 2)~Exploitation phase to exploit $K$ best bands. The entire time~horizon, $ T $ is divided into $\vartheta$ number of blocks each of duration $ 2\lceil N/K \rceil$~time slots. Depending on the value of exploration coefficient, $ L $, the algorithm explores frequency bands with the probability $ \epsilon$ and exploits with the probability $ (1-\epsilon)$ as shown in Algorithm 1 (line 6). Higher the value of $L$, higher is the number of times each band is explored. 
The value of $L$ depends on $\mu$, i.e. the minimum separation between spectrum statistics of frequency bands and is determined empirically. 

In the exploration phase (line 6-13), the LDM algorithm learns the transition probability, $ p_{uv} $ by sequentially selecting $K$ bands for two consecutive time slots. In each time slot, SNS and RS blocks digitize and sense $A_N$ bands. Since, $|A_N|=K$, $\xi_{A_N}$ becomes 0 due to which throughput, $R_{A_N}$ (i.e. a number of vacant bands in $A_N$)~is~calculated as $||\textbf{s}_{A_N}||_0$ (line 11). Let $ C^n_{uv} $ denotes the observed number of state transitions from $ u $ to $ v $ state for the $ n^{th} $ band. Then, $ \hat{p}_{uv}^n $ and $ \mathbf{\Omega}(t_s) $ are calculated using Eq.~\ref{puv} and Eq.~\ref{B_up}, respectively. 
% and $ \hat{p}_{uv}^n $ as
\vspace{-0.15em}
\begin{equation}
% C_{uv}^n, \hat{p}_{uv}^n 
\hat{p}_{uv}^n = \frac{C_{uv}^n}{C_{uv}^n + C_{uu}^n}
\label{puv}
\vspace{-0.25em}
\end{equation}
In the exploitation phase (line 15-20), the LDM~algorithm~selects $K$ best quality frequency bands by maximizing the expected immediate throughput as shown in line 16 where probability of successful reconstruction, $\mathbb{P}(\xi_{A_N}=0) = 1$. Similar~to~the~exploration~phase, SNS and RS blocks digitize and sense the selected bands followed by the calculation of $R_{A_N}$ (line 18) and $\mathbf{\Omega}(t_s)$ (Eq.~\ref{B_up}), respectively.
\setlength{\textfloatsep}{1pt}

\begin{algorithm}[!t]
	
	\caption{LDM algorithm}
	
	\begin{algorithmic}[1]
		
		\State Input: $ N, K, T $
		\State Parameter: $L$, $\mathcal{A}=$ All possible sets of $A_N$
		\State Initialization: Set~$t_s = 0$,~$ \vartheta = \lceil \frac{T}{2\lceil N/K\rceil}\rceil $, 
		$ \mathbf{\Omega}(0) = [0.5]_{N\times 1} $  $C_{uv}^{n} = 1,~\hat{p}_{uv}^n = 0.5$ $\forall~u,v\in\{0,1\}$\ and $n \in \{1,2,..,N\}$	
		\For{$b=1 \dots \vartheta$}
		\State $ \epsilon = \min\{1,\frac{L}{b}\} $
		\If{(rand $ < \epsilon $)} \Comment{Explore}
	%	\State			Explore:
		\For{$ l = 1,2,...,\lceil\frac{N}{K}\rceil $}	 
		\State 				$ A_N = \{(l-1)K+1,...,\min(lK,N)\}$
		\For{$ q = 1,2$}
		\State				Perform SNS and RS to determine $ \textbf{s}_{A_N} $
		\State				$R_{A_N}(t_s) = ||\textbf{s}_{A_N}||_0(1-\xi_{A_N})$
	%	\State				$ t_s = t_s+1 $ %and
		%		\State				Update $ \mathbf{\Omega}(t_s) $
		\EndFor
		
		\EndFor
		\Else \Comment{Exploit}
	%	\State			:
		\For{$ l = 1,2,...,2\lceil\frac{N}{K}\rceil $}
		\State			    \small{$A_N=\argmax_{A_N'\in\mathcal{A}}\mathbb{P}(\xi_{A_N'}=0)\sum_{n\in A_N'}\omega_n(t_s)$}
		\State				Perform SNS and RS to find $ \textbf{s}_{A_N} $
		\State				$R_{A_N}(t_s) = ||\textbf{s}_{A_N}||_0(1-\xi_{A_N})$
	%	\State				$ t_s = t_s+1 $ and 	Update $ \mathbf{\Omega}(t_s) $
		\EndFor
		\EndIf	
		\State			Update $ C_{uv}^n, \hat{p}_{uv}^n $ and $ \mathbf{\Omega}(t_s)$. Increment $ t_s$
		\EndFor
	\end{algorithmic}
\end{algorithm}

\subsubsection{Optimized LDM Using Online Learning}
The LDM algorithm assumes $|A_N|= K$. However, by exploiting the spectrum sparsity, more than $K$ number of bands can be sensed which may offer higher throughput. 
The optimum value of $|A_N|$ can be determined by maximizing the average throughput~as
%solving the following unconstrained optimization problem which maximizes the average throughput
\vspace{-0.5em}
\begin{equation}
\label{tp}
|A_N|~=~\argmax_{|A'_N|\geq K}~\mathbb{P}(\xi_{A_N'}=0)\sum_{n\in A_N'}\hat{p}_0^n 
\vspace{-.75em}
\end{equation} 
where $ \hat{p}_0^n = \frac{\hat{p}_{10}^n}{(\hat{p}_{10}^n+\hat{p}_{01}^n)}$ is an estimated vacancy probability of $n^{th}$ band in $A_N$ 
and $\mathbb{P}(\xi_{A_N}=0)$ is the probability of successful reconstruction for a given $A_N$ and is given by 
\vspace{-0.15em}
\begin{equation}
\label{prob_rec_fail}
{\mathbb{P}(\xi_{A_N}=0) = \begin{cases} 1 & \mbox{if}~|A_N|\leq K\\
\sum_{i=1}^{\big\lfloor\frac{K}{2}\big\rfloor}\mathbb{P}(||s_{A_N}||_0=i) & \mbox{if}~K<|A_N|<N
\end{cases}	}
	 \vspace{-0.25em}
\end{equation} 
With $|A_N|=K$, LDM has $\mathbb{P}(\xi_{A_N}=0) = 1$ but it~does~not guarantee optimum throughput. 
%	 In the optimised LDM (OLDM) algorithm, the value of $|A_N|$ is dynamically tuned to the maximum possible value by exploiting the sparsity knowledge of wideband spectrum estimated via frequency band characterisation in the exploration phase. 
By balancing the trade-off between the probability of successful reconstruction and $ |A_N| $, the optimised LDM (OLDM) algorithm dynamically tunes $ |A_N| $ to the maximum possible value by exploiting the learned sparsity of the wideband spectrum.

Similar to LDM, OLDM algorithm works in two phases: 1)~Exploration and 2)~Exploitation. Exploration phase is identical to that of LDM algorithm where $ |A_N| = K $. In the exploitation phase, if the frequency band statistics are estimated precisely, then the optimum value of $ |A_N| $ is calculated using Eq.~\ref{tp}. However, since the frequency band statistics are unknown and estimated over the time, we use Theorem 1 to determine the minimum number of time slots, $W$ required by the exploration phase to guarantee $ \mu-$correct estimation (i.e. $ |\hat{p}_0^n-p_0^n |<\mu/2 ~\forall~n\in\{1,2,..,N\}$) of frequency band statistics with a probability at least $1-\delta$. 
When the number of exploration time slots exceeds $W $, then  $|A_N|$ is calculated using Eq.~\ref{tp}. 
Thereafter, the OLDM selects frequency bands by maximizing the expected immediate throughput as shown in line 16 of Algorithm~1.

\noindent \textbf{Theorem 1:} If the minimum gap between $ p^m_0 $ and $ p^n_0 $ is $ \mu, \forall m, n \in \{1, .., N\} $ and $ m \ne n $,  then  the  exploration time slots, $W$ should be at least $\frac {4}{\mu^2 } \Big\lceil\frac{N}{K}\Big\rceil \ln\Big(\frac{2N}{\delta}\Big)$ to achieve $ \mu-$correct estimation  with the probability of  $ 1-\delta $.\\
\noindent \textbf{Proof:} 
For an event, $ J $ denoting that each band has been observed minimum $ Q $ times, we can upper bound the probability of no $ \mu-$correct estimation is achieved given $ J $ as
\vspace{-0.5em}
\begin{equation}
\label{prob_Q}
\mathbb{P}\left ( \mbox{No}~\mu-~\mbox{correct estimation} | J \right) < \delta
\vspace{-0.5em}
\end{equation}
\vspace{-0.5em}
Mathematically, it can be represented as
\vspace{-0.2em}
\begin{equation*}
\mathbb{P}\left ( \exists~n\in\{1 \cdots N\}~s.t.~|\hat{p}_0^n-p_0^n |>\frac{\mu}{2}~|~J \right)
\end{equation*}
\vspace{-1em}
\begin{equation}
\leq \sum_{n=1}^{N} \mathbb{P}\left( | \hat{p}_0^n-p_0^n | >\frac{\mu}{2}~|~J\right) \tag{By Union Bound}
\end{equation} 
\vspace{-0.5em} 
\begin{equation}
= \sum_{n=1}^{N} \sum_{q=Q}^{\infty} \mathbb{P}\left (| \hat{p}_0^n-p_0^n | >\frac{\mu}{2} \right) \mathbb{P}\left( q~\mbox{observations}|q \geq Q \right)
\vspace{-0.5em}
\end{equation}
Then by applying Hoeffding's inequality and further simplifications,
\vspace{-0.8em}
\begin{equation*}
\mathbb{P}\left ( \exists~n\in\{1 \cdots N\}~s.t.~|\hat{p}_0^n-p_0^n |>\frac{\mu}{2}~|~J \right)
\end{equation*}\vspace{-1em}
\begin{equation}
\leq \sum_{n=1}^{N} 2 \exp \left(\frac{-Q \mu^2}{2} \right) \sum_{q=Q}^{\infty} \mathbb{P} \left(  q~\mbox{observations}|q \geq Q \right) 
\end{equation}\vspace{-1em}
\begin{equation}
\label{prob_eq}
\leq 2N \exp \left(\frac{-Q \mu^2}{2} \right)
\vspace{-0.5em}
\end{equation}
From Eq.~\ref{prob_Q}, the above equation can be written as
\vspace{-0.5em}
\begin{equation}
2N \exp \left(\frac{-Q  \mu^2}{2} \right) < \delta \newline
\implies  Q >  \frac{2}{\mu^2} \ln \left ( \frac{2 N}{\delta}  \right)
\vspace{-0.2em}
\end{equation}
\noindent Since $ \mathbb{P} ($No $ \mu-$correct estimation$|J) <\delta $ implies $\mathbb{P}( \mu-$correct estimation$|J)\geq 1-\delta$, therefore $ Q $ should be greater than $  \frac{2}{\mu^2} \ln \left ( \frac{2 N}{\delta}  \right) $ for $ \mu-$correct estimation.  As in every $ 2\Big\lceil\frac{N}{K}\Big\rceil $ time slots, only one observation of each frequency band is obtained. Thus the number of time slots required to obtain $Q$ observations of all bands  (i.e. $ \mu-$correct estimation) is given by
\vspace{-0.5em}
\begin{equation}
\label{bbb}
W \geq \frac {4}{\mu^2 } \ceil[\Bigg]{\frac{N}{K}} \ln\Bigg(\frac{2N}{\delta}\bigg)
\end{equation} 
\vspace{-2.5em}
\begin{flushright}
	$\blacksquare$
\end{flushright}
\vspace*{-\baselineskip}
\vspace{-0.5em}	
\section{Performance Analysis}
\vspace{-0.2em}	  
In this section, we evaluate the performance of the proposed WSS approach on the synthetic data generated in MATLAB as well as on the NI USRP hardware testbed.
Performance of the proposed algorithm is compared with an ideal myopic policy (IMP) which has prior knowledge of spectrum statistics as well as optimum value of $|A_N|$ \cite{mab_cs}. The performance metrics used for the comparison are total average throughput and regret where regret is the difference between the average throughput of IMP and that of the proposed algorithms. We consider the signal bandwidth of 2 GHz and $N=8$ bands with two cases depicting different spectrum statistics. The optimum value of $|A_N|$ is 7 and 5 for Case 1 and 2, respectively. The respective stationary probabilities are\\
Case 1: $ \textbf{p}_0 = [0.60~0.65~0.70~0.75~0.80~0.85~0.90~0.95]$\\
Case 2: $ \textbf{p}_0 = [0.45~0.50~0.55~0.60~0.65~0.70~0.80~0.90]$ 
\subsubsection{Simulation Results}The average throughput and regret comparison of three algorithms, IMP, LDM and OLDM for Case 1 and $K=4$ are shown in Fig.~3(a). It can~be~observed that the OLDM algorithm offers better performance~than~LDM algorithm due to the estimation of optimum $|A_N|$. Furthermore, the average regret of OLDM saturates after 3,000 time slots which implies that OLDM achieves a $ \mu $-correct estimation of $ \textbf{p}_0$ and hence, its instantaneous throughput converges to that of IMP which has prior knowledge of these spectrum statistics.
The average regret observed at a signal to noise ratio (SNR) of 20~dB for two different spectrum statistics is shown in Fig.~3(b). Since the optimum value of $ |A_N| $ is 7, and 5 for Case~1, and Case~2, respectively, the average throughput achieved by IMP is higher for Case~1 and hence, the average regret during the exploration time is higher for Case 1.

The regret comparison of OLDM and LDM algorithms for different values of $ K $ and $ N $ is shown in Fig.~3(c).~It~can~be~observed that due to the requirement of lesser exploration time for larger $K$, the regret decreases with an increase in $K$. Furthermore, the regret increases with $N$ due to the requirement of higher exploration time for the larger value of $N$. Same results can be verified from Theorem~1. The average throughput for a wide range of SNRs with $N=8$, $K=4$, $T=10,000$ and spectrum statistics of Case~1 is shown in Fig.~3(d). As expected, due to the improved performance of the RS block, the average throughput increases with an increase in SNR and OLDM offers consistently better performance than LDM.

\subsubsection{Experimental Results}	
In the proposed testbed,  USRP-2922 with VERT900 antennas are used for the wireless transmission and reception of the multiband signal. The baseband signal processing is done in the LabVIEW environment. Parameters used for the transmitter and receiver model are IQ sampling rate~=~500~ksps, carrier frequency~=~935~MHz and RF receive/transmit gain of 6~dB.
Fig.~4 compares the average throughput and regret for Case~1 and Case~2. Similar to the simulation results, due to the higher optimum value of $ |A_N| $ for Case 1, the total average throughput for Case 1 is higher than that of Case 2. It can also be observed that regret of the proposed approach is constant after $ 3,000 $ time slots which implies that OLDM achieves a $ \mu $-correct estimation of $ \textbf{p}_0$ and hence, optimum $ |A_N| $ in real radio environment as well.

%that the instantaneous throughput of the proposed OLDM algorithm becomes same as that of IMP. This confirms that the proposed online learning algorithm estimates frequency band statistics and determines optimum $ |A_N| $ value in the real radio environment.
\begin{figure}[!t]
	\label{P_form}	
	\vspace{-1em}
	\centering
	\subfloat[]{\includegraphics[width=4.4cm,height=3cm]{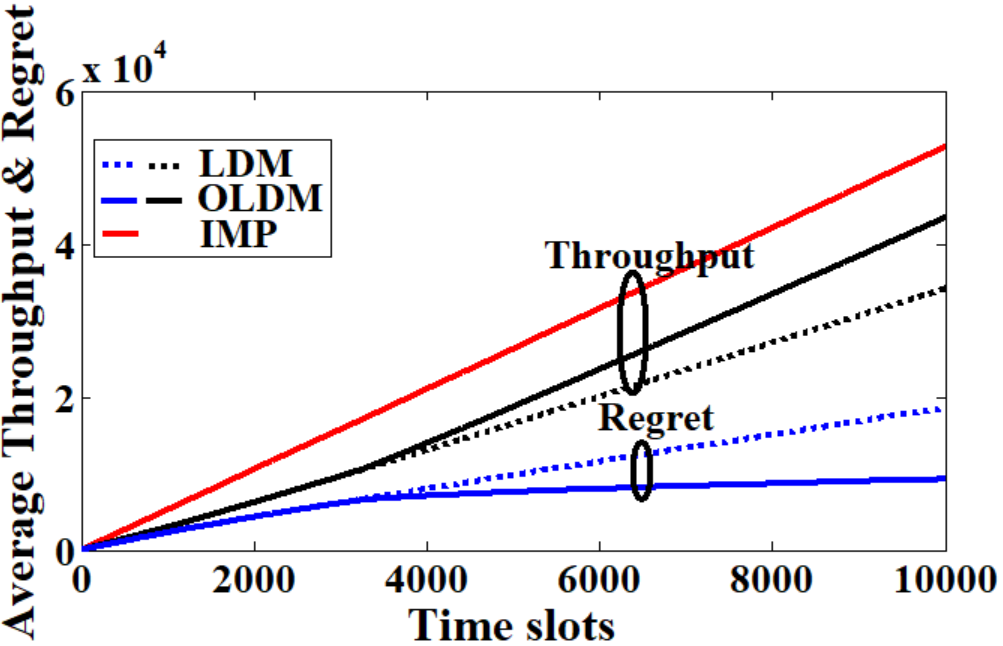}}
				  	  	  	\hspace{0.03cm}
	\subfloat[]{\includegraphics[width=4.3cm,height=2.8cm]{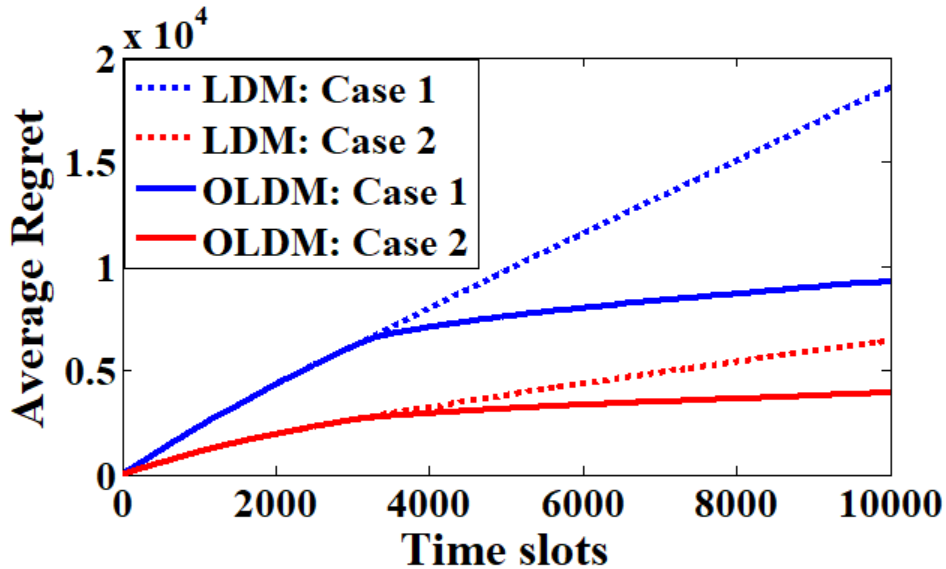}}\\
	\vspace{-1em}
	\subfloat[]{\includegraphics[width=4.3cm,height=2.8cm]{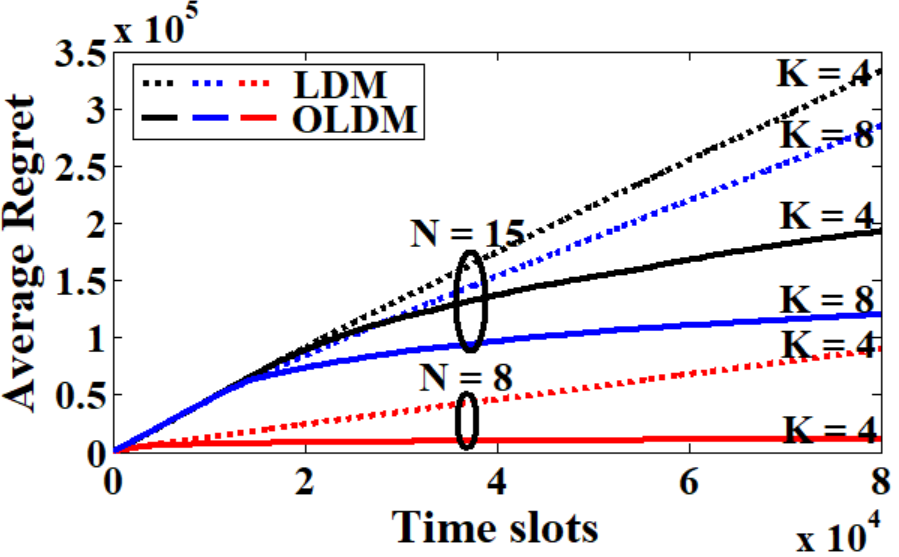}}
	\hspace{0.08cm}
	\subfloat[]{\includegraphics[width=4.3cm,height=2.75cm]{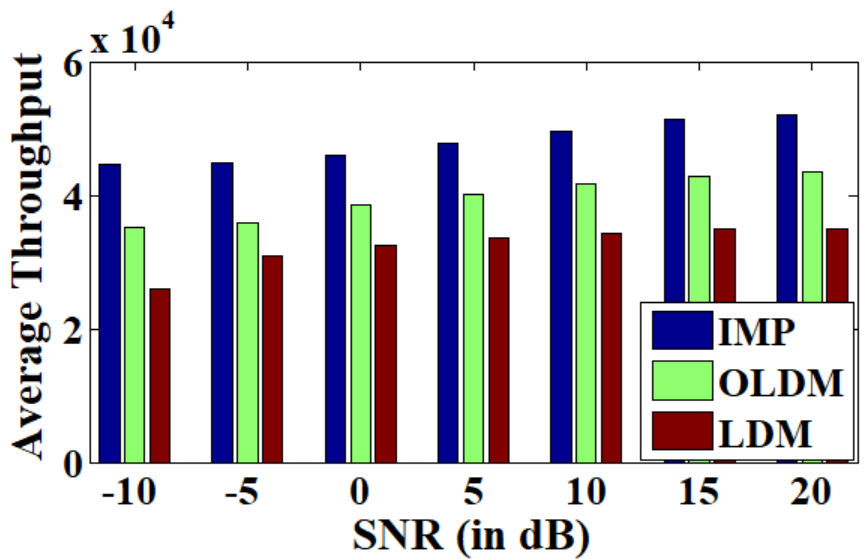}}
	\vspace{-0.5em}
	\caption{Simulation results for (a)~Average throughput and regret of LDM and OLDM w.r.t IMP (b)~Average throughput for three different sets of spectrum statistics (c)~Average regret for different values of K and N (d)~Comparison of average throughput for LDM, OLDM and IMP for different values of SNRs }
			\vspace{-0.5em}
	%	\end{figure}
	%	\begin{figure}[!h]
	\label{P_form}	
	\vspace{-0.5em}
	\centering
	\subfloat[]{\includegraphics[width=4.3cm,height=2.75cm]{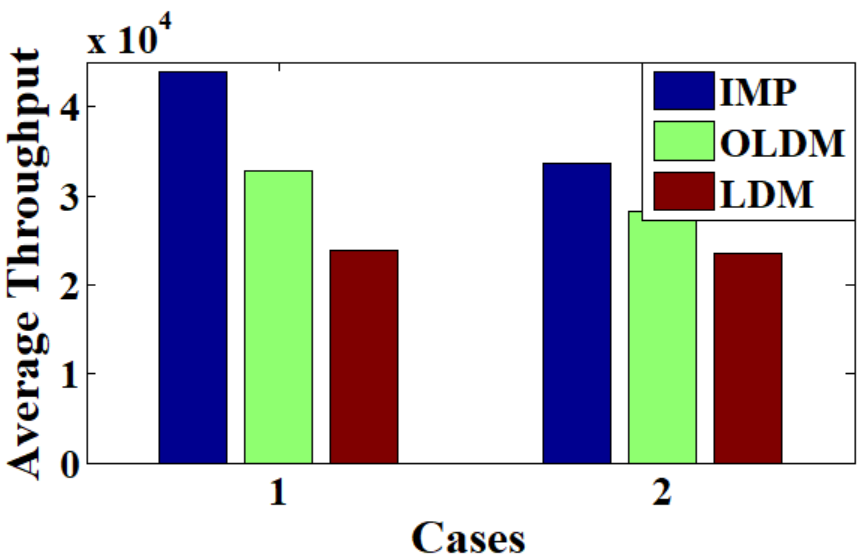}}
	\hspace{0.085cm}
	\subfloat[]{\includegraphics[width=4.3cm,height=2.8cm]{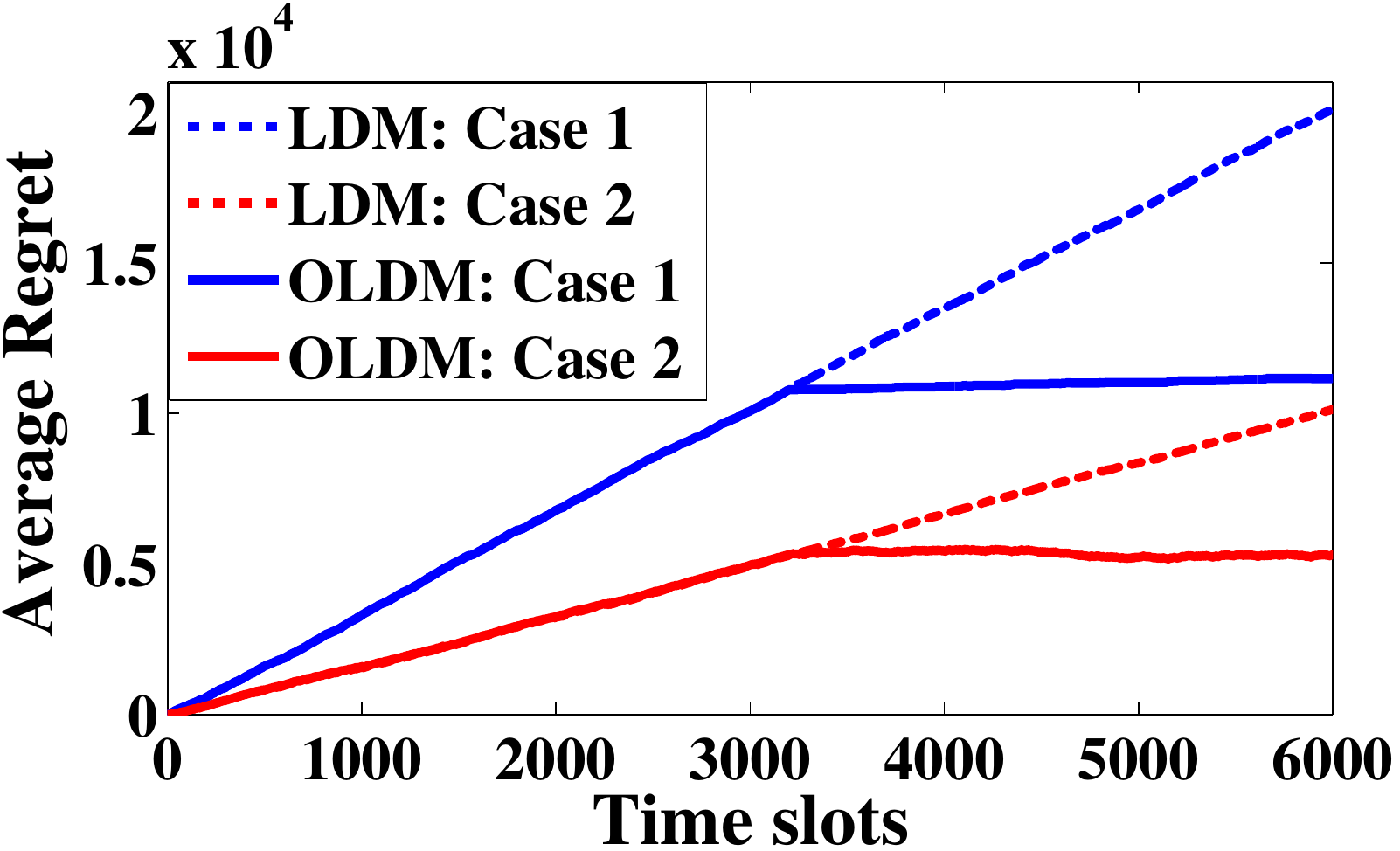}}
	\vspace{-0.5em}
	\caption{Experimental results for (a) Average throughput for Case 1 and Case 2 (b) Average regret for Case 1 and Case 2 }
%	\vspace{-0.1em}
\end{figure}
%\vspace{-0.5em}
\section{Conclusions}
%	\vspace{-0.1em}
In this brief, we proposed a non-contiguous WSS approach using sub-Nyquist sampling and novel online learning algorithm to characterize and select frequency bands based on their spectrum statistics. Theoretical guarantees, extensive simulation and experimental results in the real radio environment validate the superiority of the proposed approach. Future works include an extension of the proposed approach for WSS in the spatial domain.	

\vspace{-0.6em}
%   \section{Acknowledgements}
%	This work is supported by the funding received from Council of Scientific and Industrial Research (CSIR), India under Junior Research Fellowship (JRF) Scheme and Department of Science \& Technology (DST), India under Innovation in Science Pursuit for Inspired Research (INSPIRE) faculty fellowship.

%	\vspace{-0.5em}
			
\end{document}